\documentclass[aps,pre,twocolumn,showpacs]{revtex4} 
 
\usepackage{amsfonts} 
\usepackage{amssymb} 
\usepackage{graphicx}

\newcommand{\eqref}[1]{Eq.~(\ref{#1})}

\begin{document} 
 
\title{Nonlocal incoherent solitons} 
 
\author{Wies{\l}aw Kr\'olikowski} 
 
\affiliation{Laser Physics Centre,  
Research School of Physical Science and Engineering, \\ 
The Australian National University, Canberra ACT 0200, Australia} 
 
\author{Ole Bang} 
 
\affiliation{Research Center COM, Technical University of Denmark,  
         2800 Kongens Lyngby, Denmark} 
 
\author{John Wyller\footnote{On sabbatical leave from the Department of 
Mathematical Sciences, Agricultural University of Norway, P.O.Box 5065, 
N-1432 \AA s, Norway.} 
} 
 
\affiliation{Nonlinear Physics Group and Laser Physics Centre,  
Research School of Physical Science and Engineering, \\ 
The Australian National University, Canberra ACT 0200, Australia} 

\begin{abstract} 
We investigate the propagation of partially coherent beams in  
spatially nonlocal nonlinear media with a logarithmic type of  
nonlinearity.    
We derive analytical formulas for the evolution of the beam  
parameters and conditions for the formation of nonlocal  
incoherent solitons. 
\end{abstract} 
 
\pacs{42.65Tg, 42.65.Jx, 42.25.Kb}
\maketitle 
 
Incoherent solitons are partially coherent optical beams propagating  
without changing their shape in materials with a slow nonlinear response 
\cite{incoh1,incoh2}.  
Such nonlinear materials respond to the time averaged light intensity,  
which means that even though the phase of the partially coherent light  
beam fluctuates randomly, it still leads to a smooth light-induced  
refractive index change and subsequent trapping of the beam
\cite{mm1,csms,csj,csms2,newstuff,big,shkunov}.
Incoherent solitons have been experimentally observed almost solely  
in photorefractive media, which exhibit a strong and sufficiently slow 
saturable Kerr-like nonlinearity \cite{incoh1,incoh2}.  

In inertial bulk Kerr media (2+1)-dimensional partially coherent beams 
are still unstable and will either diffract or self-focus unboundedly, 
depending on whether the power is above a certain critical value 
\cite{us_incor_coll,Aleshkevich,Pasmanik}. Incoherence counteracts
self-focusing and leads to an increase of the critical power, but it 
cannot remove the collapse instability \cite{us_incor_coll,Aleshkevich}.
Consequently all experimental observations of incoherent solitons have
been in saturable materials, such as photorefractive materials 
\cite{incoh1} and liquid crystals \cite{lc_incoherent}.

Recently Peccanti {\em et al.} reported the first observation of incoherent
solitons and their interaction in nematic liquid crystals 
\cite{lc_incoherent}.  
Liquid crystals have a strong non-instantaneous saturable Kerr-like
nonlinearity associated  with light induced molecular reorientation 
\cite{lc1,lc2,lc3,lc4}. 
Their nonlinear response time ranges from tens to hundreds of  
milliseconds \cite{lc1}, which is sufficiently slow to allow for the  
formation of incoherent solitons.  
Apart from being inertial the nonlinear response of liquid crystals is  
also inherently {\em spatially nonlocal}, because the molecular  
reorientation induced by a light beam in a particular place will affect  
the orientation of molecules far beyond this point, due to the long-range  
character of their interaction \cite{lc2,lc_waveguide}.  
 
Spatially nonlocal nonlinearities are in fact an inherent property of  
many physical systems, including matter waves, nonlinear optics, and  
field theory. 
As such the nonlocal aspect of nonlinear interaction has attracted  
significant interest recently \cite{rev1,rev2}.  
Nonlocality can have profound consequences on the properties of optical beams and  
soliton formation, e.g., leading to collapse arrest of finite-size beams 
\cite{us_collapse}, attraction and formation of bound states of dark solitons 
\cite{us_darkbound}, and modulational instability (MI) in defocusing  
materials \cite{us_MI,us_MI_generic}. 
A nonlocal nonlinearity may even describe parametric wave mixing and solitons 
\cite{us_chi2nonlocal}. 
In the particular case of liquid crystals certain aspects of their  
nonlocal nonlinear response have already been discussed in the literature,  
including theoretical and experimental studies of the formation of  
multiple solitons \cite{Hutsebaut_04,McLaughlin}, accessible solitons \cite{lc_accessible} as well as soliton  
interaction and MI \cite{lc_MI}. 
 
Motivated by the documented unique dual inertial and nonlocal character  
of the nonlinearity of liquid crystals and the recent experimental 
observation of incoherent solitons in this material, we analyze here 
the effect of nonlocality on the propagation of partially coherent 
beams and formation of incoherent solitons.  
Because such a problem is in general analytically intractable we consider here  
the special case of a nonlinearity with a logarithmic dependence of   
the refractive index change on light intensity.   
This model has previously been used to study incoherent beams in local  
media \cite{csj,csms2,mighty,KrolEdmundBang00,Ponomarenko} as well as fully 
coherent beams in nonlocal media \cite{accessible}.  
Here we extend it to describe {\it simultaneously incoherence and nonlocality}. 
While being rather specific, the logarithmic model is unique in the extent  
that it enables a fully analytical treatment of the nonlinear beam  
propagation. It has been successfully used to predict a variety of effects  
associated with beam propagation and soliton formation  
\cite{KrolEdmundBang00,gaus}.    
 
Propagation of 
a two-dimensional quasi-monochromatic partially coherent beam with  
the slowly varying amplitude $\psi(\vec{r},z)$ is governed by the  
following nonlinear Schr\"{o}dinger equation: 
\begin{equation}\label{nls1} 
i\frac{\partial\psi}{\partial z} + \frac{1}{2}\nabla^2_{\vec{r}}\,\psi 
        + \delta n(I) \psi = 0, 
\end{equation} 
where $\vec{r}$=$(x,y)$, $I$=$|\psi|^2$ denotes the light intensity, 
 $\delta n(I)$ represents the light induced refractive index change, 
and $\nabla^2_{\vec{r}}=\partial^2/\partial x^2 + \partial^2/\partial y^2$.  
 
In the following we use the mutual coherence function   
$\Gamma(\vec{r}_1,\vec{r}_2,z)$ to represent the coherence properties  
of the beam \cite{Wolf}.  
This function describes the spatial correlation between the field  
at points $\vec{r}_1$ and $\vec{r}_2$, and is defined as  
\begin{equation} 
\Gamma(\vec{r}_1,\vec{r}_2,z) = \langle\psi(\vec{r}_1,z) 
                              \psi^*(\vec{r}_2,z)\rangle, 
\end{equation} 
where brackets denote temporal or ensemble averaging.   
In terms of the mutual coherence function the (time averaged)
intensity is given by 
\begin{equation}
   I(\vec{r},z)=\Gamma(\vec{r},\vec{r},z).
\end{equation}
We take the refractive index change $\delta n$ to be the following   
{\em nonlocal} function of light intensity 
\begin{equation}\label{deltan} 
   \delta n(\vec{r},I) = n_2\ln \left[\int R(\vec{r}-\vec{\xi}\,)
                         I(\vec{\xi}\,)\, d\vec{\xi}\right]. 
\end{equation}  
where $\int d\vec{r}=\int_{-\infty}^{\infty}\int_{-\infty}^{\infty} dxdy$ 
and $R(\vec{r})$=$R(r)$ is the so-called nonlocal response function,  
which depends only on the length of its argument vector, $r=|\vec{r}|
=\sqrt{x^2+y^2}$.  

The model (\ref{deltan}) is phenomenological and reflects the fact that  
the refractive index change in a particular point in space is  
not only determined by the wave intensity in this point, but also on 
the intensity in a certain surrounding region given by the width of 
the response function. 
The width of the nonlocal response function therefore determines the  
degree of nonlocality.  
In the special case when $R(\vec{r})=\delta(r)$ this model describes a local  
nonlinear medium $\delta n=n_2\ln I$. 
A simplified model may also be derived in the weakly nonlocal limit,
in which the width of the response function is finite, but small 
compared to the beam width \cite{us_weakly,rev2}. 
In the strongly nonlocal limit, in which the response function is much
broader than the beam, the governing model becomes linear 
\cite{accessible,rev2}.  

  In general, the particular form of the response function is determined  
by the specifics of the physical process responsible for the nonlinearity  
of the optical medium. For instance, it can be shown that for the  
reorientational nonlinearity of liquid crystals and general diffusion  
type nonlinearities, the nonlocal response can be approximated as  
\begin{equation} 
   R(\vec{r}) \propto\exp(-r/\sigma), 
\end{equation} 
with $\sigma$ representing the extent of the nonlocality.  
However, the majority of nonlocality-mediated effects appear to be rather   
generic \cite{rev2,us_MI_generic} and do not depend strongly on the particular  
form of the nonlocal response function.  
Here, to enable us to make an analytical description, we therefore assume  
that the nonlocality is described by the normalized Gaussian response function 
\begin{equation} 
  R(r)= \frac{1}{\pi\sigma^2}\exp\left(-\frac{x^2+y^2}{\sigma^2} \right ). 
\end{equation} 
While we consider here a nonlocally isotropic medium, the model can be 
easily extended to include anisotropy of the nonlocal properties, such  
as those caused by a directional heat flow in thermal media \cite{thermal}. 
  
The mutual coherence function satisfies the following propagation equation   
\cite{Pasmanik,Aleshkevich} 
\begin{eqnarray} 
i\frac{\partial \Gamma}{\partial z}  
   &+& \frac{1}{2}\left (\nabla^2_{\vec{r}_1} 
       - \nabla^2_{\vec{r}_2} \right )\Gamma \nonumber \\  
   &+& [\delta n(\vec{r}_1,I)-\delta n(\vec{r}_2,I)]\Gamma = 0, 
\end{eqnarray} 
Introducing the independent spatial variables $\vec{p}$ and $\vec{q}$, 
\begin{equation}\label{variables} 
   \vec{p} = \frac{1}{2}(\vec{r}_1+\vec{r}_2),  \,\,\,  
   \vec{q} = \vec{r}_1-\vec{r}_2, 
\end{equation} 
and using the specific form (\ref{deltan}) of the nonlinearity 
transforms the propagation equation into 
\begin{eqnarray}\label{gamma_eq} 
& &i\frac{\partial \Gamma}{\partial z} + 
    \nabla_{\vec{p}}\cdot\nabla_{\vec{q}}\,\Gamma + \nonumber \\ 
& &    n_2\ln\left(\frac{\int R(\vec{r}_1-\vec{\xi}\,)I(\vec{\xi}\,)d\vec{\xi}} 
{\int R(\vec{r}_2-\vec{\xi}\,)I(\vec{\xi}\,)d\vec{\xi}}\right)\Gamma = 0. 
\end{eqnarray} 

In what follows we will consider the so-called Gaussian-Schell model for the 
incident partially coherent beam \cite{Wolf}. 
We will also restrict our analysis to circular beams, i.e. beams having 
isotropic coherence properties.  
The extension of the analysis to elliptical incoherent beams 
is straightforward.     
The mutual coherence function of the Gaussian-Schell beam is expressed as  
\begin{equation}
   \label{cofun} 
   \Gamma(\vec{p},\vec{q},z=0) = \exp\left(- \frac{|\vec{p}|^2}{\rho_0^2 }- 
   \frac{|\vec{q}|^2}{\theta_{0}^2} \right), 
\end{equation} 
where the initial effective coherence radius $\theta_0$ is given 
by the relation
\begin{equation}
   1/\theta_{0}^2=1/r_{c}^2+1/(4\rho_{0}^2).
\end{equation}
Here $\rho_0$ is the initial radius and $r_c$ is the initial 
coherence radius of the beam, respectively.  
Due to the logarithmic form of the nonlinearity, the beam will maintain
the Gaussian statistics during propagation and thus the coherence function
will keep the form given by Eq.~(\ref{cofun})
We therefore look for the solutions to \eqref{gamma_eq} using the 
following Gaussian ansatz 
\begin{eqnarray} 
& &\Gamma(\vec{p},\vec{q},z)= \nonumber \\ 
& & A(z)\exp\left (- \frac{|\vec{p}|^2}{\rho^2(z)}- 
\frac{|\vec{q}|^2}{\theta^2(z)}  
    + i\vec{p}\cdot\vec{q}\mu(z)\right), \label{Gamma} 
\end{eqnarray} 
where $A(z)$ and $\mu(z)$ represent the amplitude and phase variations 
of the coherence function, and $\rho(z)$ and $\theta (z)$ are  its diameter 
and generalized coherence radius, respectively.  
The input amplitude plays no role for the dynamics and thus we use the
initial conditions 
$A(0)$=1, $\rho(0)$=$\rho_0$, and $\theta(0)$=$\theta_{0}$, 
Inserting the ansatz (\ref{Gamma}) into \eqref{gamma_eq} leads to a 
set of ordinary 
differential equations for the parameters of the coherence function 
\begin{eqnarray} 
   \frac{d\theta}{dz}  & = & \theta\mu, \label{teta}\\  
    \frac{d\rho}{dz}   & = & \rho\mu, \label{rho}\\  
    \frac{dA}{dz}      & = & -2A\mu, \label{A}\\  
    \frac{d\mu}{dz}    & = & \frac{4}{\theta^2\rho^2} -      
    \mu^2 - \frac{2n_2}{\rho^2+\sigma^2}. \label{mu} 
 \end{eqnarray} 
 
From the first two equations (\ref{teta}) and (\ref{rho}) we obtain the 
relation
\begin{equation}\label{speckle}
  \theta/\rho = \theta_0/\rho_0,
\end{equation} 
which shows that during the evolution the beam conserves its coherence,
defined as the number of speckles within the beam diameter.
Interestingly, this relation does not depend on whether the nonlinearity 
is local or not. 
%It is a property specific to the logarithmic model of the medium. 
%which shows that during evolution the beam conserves its coherence, 
%defined as the number of speckles within the beam diameter.  The wider 
%(narrower) the beam the larger (smaller) the coherence radius. 
%Combining \eqref{rho} and \eqref{A} gives the amplitude $A(z)=[\rho_0 
%/\rho(z)]^2$.  
Combining   Eq.(\ref{rho}), Eq.(\ref{mu}) and Eq.(\ref{speckle}) we 
obtain the following evolution equation for the beam radius 
\begin{equation}\label{dynamics} 
\frac{d^2\rho}{dz^2} - 4\frac{\rho_{0}^2}{\rho^3\theta_{0}^2} 
   + 2n_2\frac{\rho}{\rho^2+\sigma^2} = 0. 
\end{equation} 
This equation describes the dynamics of the width of a partially coherent 
beam with Gaussian statistics in a nonlocal medium with a logarithmic 
nonlinear response. 
Taking  $\sigma=0$ we recover the expressions governing incoherent beams 
in a local medium \cite{KrolEdmundBang00}. 
On the other hand, for $r_{c}\rightarrow\infty$, Eq.~(\ref{dynamics}) 
describes coherent beams in a nonlocal medium.  It is 
clear that the main role of the nonlocality is to effectively weaken the 
influence of the nonlinearity on the propagation of the beam. For very high 
degrees of nonlocality the nonlinear effects becomes negligible and the beam 
just diffracts.  
 
Incoherent nonlocal soliton solutions are obtained from Eq.~(\ref{dynamics}) 
by setting $\rho(z)=\rho_0$. This gives the relation
\begin{equation} \label{n_2} 
  n_2 = \left( \frac{2}{r_c^2}+\frac{1}{2\rho_0^2}\right)
        \left(1+\frac{\sigma^2}{\rho_0^2}\right).
\end{equation} 
As expected, the existence of bright incoherent solitons, or localized 
beams, is seen to require a focusing nonlocal nonlinearity with $n_2>0$.
Importantly, the relation (\ref{n_2}) further shows that trapping of a 
beam with a given initial width $\rho_0$ and coherence radius $r_c$ in 
a nonlocal medium requires a {\em stronger nonlinearity} (higher $n_2$)
than in the case of a purely local nonlinear response. 
This is again due to the fact that the nonlocality effectively leads 
to a decrease of the strength of focusing and subsequently to a weaker  
localisation of the beam. From Eq.~(\ref{n_2}), we find the expression 
for the radius  of the incoherent nonlocal soliton  
\begin{equation} 
  \rho_0^2 =
  \frac{\tilde{\rho_0}^2}{2}\left[ 1+\frac{4\sigma^2}{r_c^2}+
      \sqrt{\left(1+\frac{4\sigma^2}{r_{c}^2}\right)^2+\frac 
      {4\sigma^2}{\tilde{\rho_{0}}^2}}~\right], 
\end{equation} 
where 
\begin{equation}
  \tilde{\rho_0}^2 \equiv \rho_0(\sigma=0) = \frac{1}{2n_2-4/r_c^2}
\end{equation}
denotes the radius of the incoherent soliton in a local medium. 
 
\begin{figure}[h] 
  \centerline{\scalebox{0.45}{\includegraphics{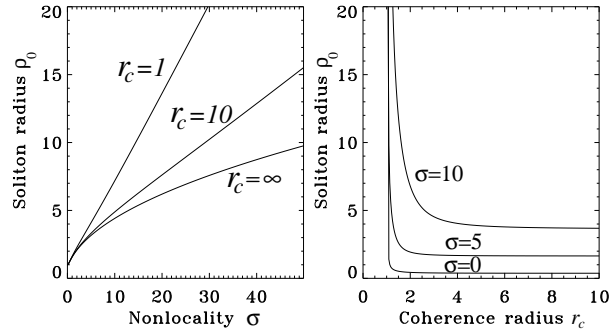}}} 
  \caption{(a) Soliton width ($\rho$) as a function of the degree of  
  nonlocality (a) and coherence radius (b). Here $n_2$=1.8} 
  \label{fig1} 
\end{figure} 
 
In Fig.~\ref{fig1}(a) we plot the radius of incoherent nonlocal solitons
versus the degree of nonlocality $\sigma$ for three values of 
the coherence radius $r_c$. 
These plots clearly show that the soliton width increases with
nonlocality and becomes proportional $\sigma$ in the highly nonlocal 
regime. One can show that for $\sigma/r_c\gg1$ 
\begin{equation} 
  \rho_0 = 2\tilde{\rho_0}\frac{\sigma}{r_c}. 
\end{equation}  
The degree of nonlocality is an inherent property of the medium and as 
such difficult to control. However, one can easily vary the coherence 
properties of an incident beam. Therefore we plot in Fig.~\ref{fig1}(b) 
the soliton width $\rho_0$ versus the coherence radius $r_c$ for three 
values of the nonlocality $\sigma$. 
Again, the increase of the soliton width with incoherence and nonlocality 
is evident. Additionally, Fig.~\ref{fig1}(b) shows that for a given 
strength of nonlinearity there is minimum coherence radius, below 
which the incoherent soliton cannot exist. 
The existence of such a threshold is well-known from earlier works on 
incoherent solitons in local media \cite{csj,csms2,KrolEdmundBang00}.  
Interestingly, it turns out that for a nonlocal nonlinearity, this 
threshold is also determined solely by the strength of the nonlinearity 
and the degree of coherence through the relation
\begin{equation}  
  n_2-2/r_c^2>0. 
\end{equation} 
To understand why this condition does not depend on the beam size or 
the degree of nonlocality we recall that the soliton is formed when 
the nonlinearity fully compensates the transverse spreading of the 
beam, which is caused by incoherence and diffraction. 
The former is due to the diffusive character of the beam, while 
the latter is a direct consequence of the finite size of the beam. 
For a very broad beam the contribution to its expansion due to 
diffraction and nonlocality becomes negligible compared to that 
due to incoherence. 
Hence, to achieve beam trapping the strength of the nonlinear response 
of the medium should be high enough to compensate at least the 
incoherence-induced beam spreading. 
This, in fact, determines the threshold for soliton generation. 
As soon as this condition is satisfied a bright incoherent soliton
can exist, whose width will be determined by the interplay of 
nonlinearity, coherence, diffraction, and nonlocality. 
Obviously, the resulting beam diameter will always be larger than 
that of the local soliton of the same degree of coherence.  
 
\begin{figure}[h] 
\centerline{\scalebox{0.45}{\includegraphics{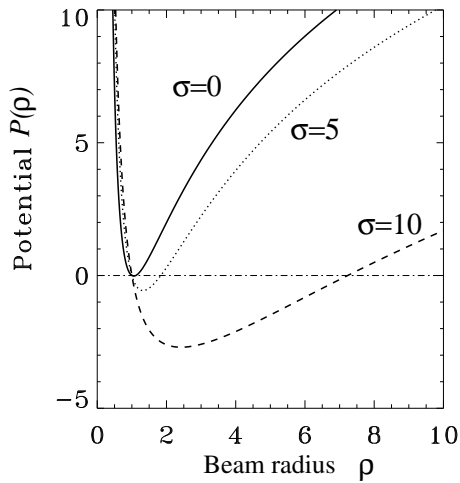}}} 
\caption{(a) Potential $P(\rho)$ in a focusing logarithmic nonlinear  
medium ($n_2$=1.8) for different degrees of nonlocality $\sigma$.  
The initial beam parameters are $\theta_0$=1 and $\rho_0=1$.} 
\label{figpot} 
\end{figure}

More physical insight into the competition between diffraction, 
nonlinearity, nonlocality, and incoherence can be obtained by an 
effective particle analogy. 
Assuming that $(d\rho/dz)(z$=$0)$=0, and integrating \eqref{dynamics} 
once, one obtains a classical-mechanical equation   
\begin{equation} 
   (d\rho/dz)^2 + P(\rho) = 0,  
\end{equation} 
describing an effective particle $\rho$ with kinetic energy $(d\rho/dz)^2$
moving in the potential $P(\rho)$, given by 
\begin{equation}\label{eqpot} 
   P(\rho) = \frac{4}{\theta_0^2}\left(\frac{\rho_0^2}{\rho^2}-1 \right) 
             + 2n_2\ln\left(\frac{\sigma^2+\rho^2}{\sigma^2+\rho_0^2}\right). 
\end{equation} 
The asymmetric potential is depicted in Fig.~\ref{figpot} for a beam 
with effective coherence radius $\theta_0$=1 moving in a focusing 
logarithmic nonlinear medium with $n_2$=1.8.  
As nonlocality increases the width of the potential also increases and 
its minimum becomes less pronounced. 
 
Stationary soliton solutions with constant beam width correspond to the 
effective particle being located at the bottom of the potential well.  
For initial beam parameters different from the soliton solution the beam 
width (and coherence radius and peak intensity) will undergo periodic 
oscillations, corresponding to the effective particle oscillating in 
the bottom of the potential well. 
As the profile of the potential well changes with higher 
$\sigma$ one can expect that the amplitude and period of beam oscillations  will 
drastically increase for high degree of nonlocality. In the limit of the small 
amplitude oscillations, their form can be obtained analytically in the form of 
Jacobi elliptical function, in the way analogous to that discussed in 
\cite{KrolEdmundBang00}. 
 
\noindent 
    
\begin{figure}[h] 
\centerline{\scalebox{0.45}{\includegraphics{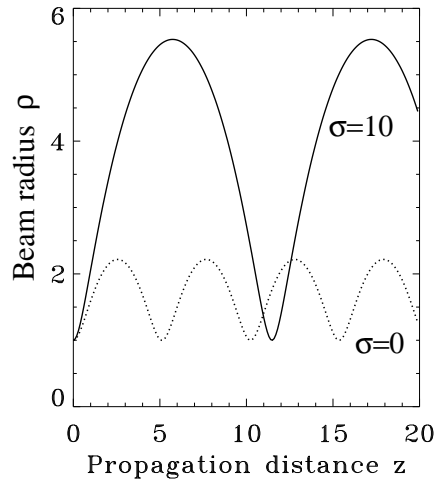}}} 
\caption{Illustration of nonstationary propagation of a partially coherent  
beam in a logarithmic nonlinear nonlocal medium. The nonlocality parameter  
varies as $\sigma=0$ (dotted line) and $\sigma=10$ (solid line).}  
\label{figdyn} 
\end{figure}

This is indeed the case as illustrated in  Fig.~\ref{figdyn}  where  we show the 
nonstationary propagation of the partially coherent beam in nonlinear medium. 
The initial parameters are chosen such that $\rho_{0}=1.0$, 
$r_{c}= 1.15$, and $n_2=1$. In the graph we plot the beam 
radius as functions of the propagation distance in case of  local (dotted line) 
and nonlocal (solid line) nonlinearity. The increased amplitude and period of 
oscillations induced by the nonlocality is evident.  
  
In conclusion, we analyzed the  
propagation of incoherent optical  beams with Gaussian statistics in a 
nonlinear nonlocal medium.  Considering a special case of  a logarithmic type of 
nonlinearity and Gaussian nonlocal response we obtained an analytical formulas 
governing all relevant parameters of the beam.  We showed  that nonlocality 
results in increased diameter of the incoherent solitons. When the initial 
parameters of the beam differ from the exact nonlocal  soliton solution, beam 
experiences periodic expansions and contractions whose  period increases with 
nonlocality. 
  
Acknowledgement: W.K. was supported by the Australian Research Council. J.W. acknowledges support from The Research Council of Norway under the grant No. 153405/432.

\end{document}